\documentclass[preprint,12pt]{elsarticle}

\usepackage{amssymb}
\usepackage{amsmath}

\journal{Physics Reports}

\usepackage[v1compatible,canvasScaleIsUnitLength]{axodraw2}

\RequirePackage{xspace}
\newcommand{\Collier}{{\sc Collier}\xspace}

\newcommand{\Recola}{{\sc Recola}\xspace}

\newcommand{\Openloops}{{\sc OpenLoops}\xspace}
\newcommand{\MGNLO}{{\sc MadGraph5\_aMC@NLO}\xspace}
\newcommand{\Gosam}{{\sc GoSam}\xspace}

\newcommand{\NLOX}{{\sc NLOX}\xspace}

\def\refeqf#1{\mbox{(\ref{#1})}}

\def\citere#1{\mbox{Ref.~\cite{#1}}}
\def\citeres#1{\mbox{Refs.~\cite{#1}}}

\newcommand{\GeV}{\unskip\,\mathrm{GeV}}
\newcommand{\MeV}{\unskip\,\mathrm{MeV}}

\newcommand{\ri}{{\mathrm{i}}}

\newcommand{\rT}{{\mathrm{T}}}

\def\mathswitchr#1{\relax\ifmmode{\mathrm{#1}}\else$\mathrm{#1}$\fi}

\newcommand{\PW}{\mathswitchr W}
\newcommand{\Pw}{\mathswitchr w}
\newcommand{\PZ}{\mathswitchr Z}

\newcommand{\Pg}{\mathswitchr g}

\newcommand{\PH}{\mathswitchr H}
\newcommand{\Pe}{\mathswitchr e}

\newcommand{\Pj}{\mathswitchr j}

\newcommand{\Pp}{\mathswitchr p}
\newcommand{\Pt}{\mathswitchr t}

\newcommand{\Pep}{\mathswitchr {e^+}}
\newcommand{\Pem}{\mathswitchr {e^-}}

\newcommand{\PWp}{\mathswitchr {W^+}}
\newcommand{\PWm}{\mathswitchr {W^-}}

\def\mathswitch#1{\relax\ifmmode#1\else$#1$\fi}

\newcommand{\MW}{\mathswitch {M_\PW}}

\newcommand{\MZ}{\mathswitch {M_\PZ}}
\newcommand{\MH}{\mathswitch {M_\PH}}

\newcommand{\Mt}{\mathswitch {m_\Pt}}

\newcommand{\rw}{\mathswitchr w}
\newcommand{\sw}{\mathswitch {s_\rw}}

\newcommand{\GF}{\mathswitch {G_\mu}}

\newcommand{\LO}{{\mathrm{LO}}}

\newcommand{\EW}{{\mathrm{EW}}}
\newcommand{\QCD}{{\mathrm{QCD}}}

\newcommand{\DPA}{{\mathrm{DPA}}}

\newcommand{\MSbar}{\overline{\mathrm{MS}}}

\newcommand{\Matrix}{{\sc Matrix}}

\newcommand{\alphas}{\alpha_\mathrm{s}}

\let\gsim\gtrsim
\let\lsim\lesssim

\usepackage{enumitem}   

\newenvironment{MYitemize}%
{\begin{itemize}[leftmargin=!,labelindent=0pt,labelwidth=2ex,%
   labelsep=1ex,topsep=2pt plus 1pt minus 1pt,itemsep=3pt plus 1pt minus 1pt,parsep=0pt]}%
{\end{itemize}}

\begin{document}

\begin{frontmatter}



\title{The status of theory in the electroweak sector: Radiative corrections, 
salient features, approximations\tnoteref{specialissue}}
\tnotetext[specialissue]{Invited contribution to the 
Special Issue ``Review of Precision Electroweak Measurements, Theory and Phenomenology''
of {\it Physics Reports}.}


\author[affi1]{Stefan Dittmaier} 

\affiliation[affi1]{organization={University of Freiburg},
            addressline={Institute of Physics, Hermann-Herder-Str.~3}, 
            city={79104~Freiburg},
            country={Germany}}

\begin{abstract}
Electroweak radiative corrections form a crucial ingredient in modern
precision calculations for particle processes at high-energy colliders
such as the Large Hadron Collider. 
The salient features of electroweak corrections as well as currently 
used techniques and concepts for their calculation are reviewed.
Recent progress in this enterprise is illustrated in a discussion of
electroweak multi-gauge-boson production processes:
massive di-boson production, vector-boson scattering, and massive tri-boson production.
\end{abstract}




\end{frontmatter}



\section{Introduction}

After about 15~years of data taking at the Large Hadron Collider (LHC),
the observed good overall agreement between measured cross sections of high-energy particle 
processes and predictions based on the Standard Model of particle physics (SM) is spectacular
(see \citeres{ATLAS:2024fkg,ATLAS:2024wla,ATLAS:2024kxj,CMS:2024gzs} and references therein).
Even in combination with results from earlier colliders or non-collider experiments,
the SM does not show any significant cracks, although we know that any solution to
open problems of present-day particle physics -- such as the clarification of the nature of
dark matter, the theoretical embedding of neutrino masses, or the explanation of the
matter--antimatter asymmetry in the Universe -- requires physics beyond the SM.
The high-luminosity phase of the LHC as well as any potential future $\Pep\Pem$
collider therefore aim at identifying signs of new physics via measurements at
the highest possible precision.

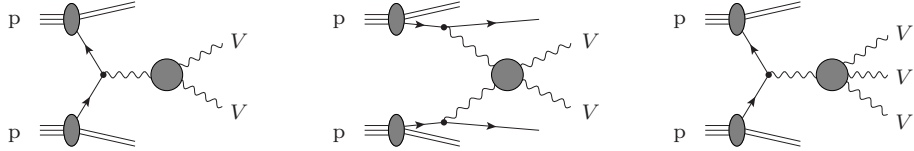
\begin{figure}
{\unitlength .6pt \scriptsize
\SetScale{0.6}
\centerline{
\begin{picture}(200,100)(-25,-10)
\Line(  0, 2)(20, 2)
\Line(  0, 5)(20, 5)
\Line(  0, 8)(20, 8)
\Line(  0,72)(20,72)
\Line(  0,75)(20,75)
\Line(  0,78)(20,78)
\Line(20, 7)(60,-2)
\Line(20, 4)(60,-5)
\Line(20,77)(60,86)
\Line(20,74)(60,83)
\ArrowLine(20, 7)(40,40)
\ArrowLine(40,40)(20,73)
\Photon(40,40)(80,40){2}{5}
\Photon(80,40)(115,60){2}{5}
\Photon(80,40)(115,20){2}{5}
\GCirc(80,40){10}{0.5}
\Vertex(40,40){2}
\GOval(20, 5)(10,5)(0){.5}
\GOval(20,75)(10,5)(0){.5}
\put(120,57){${V}$}
\put(120,12){${V}$}
\put(-20,72){$\Pp$}
\put(-20, 0){$\Pp$}
\end{picture}
\begin{picture}(210,100)(-25,-10)
\Line(  0, 2)(20, 2)
\Line(  0, 5)(20, 5)
\Line(  0, 8)(20, 8)
\Line(  0,72)(20,72)
\Line(  0,75)(20,75)
\Line(  0,78)(20,78)
\Line(20, 7)(60,-2)
\Line(20, 4)(60,-5)
\Line(20,77)(60,86)
\Line(20,74)(60,83)
\ArrowLine(20, 7)(50,10)
\ArrowLine(20,73)(50,70)
\ArrowLine(50,10)(110, 5)
\ArrowLine(50,70)(110,75)
\Photon(50,70)(90,40){2}{5}
\Photon(50,10)(90,40){2}{5}
\Photon(90,40)(130,60){2}{5}
\Photon(90,40)(130,20){2}{5}
\GCirc(90,40){10}{0.5}
\Vertex(50,70){2}
\Vertex(50,10){2}
\GOval(20, 5)(10,5)(0){.5}
\GOval(20,75)(10,5)(0){.5}
\put(137,57){${V}$}
\put(137,12){${V}$}
\put(-20,72){$\Pp$}
\put(-20, 0){$\Pp$}
\end{picture}
\begin{picture}(180,100)(-25,-10)
\Line(  0, 2)(20, 2)
\Line(  0, 5)(20, 5)
\Line(  0, 8)(20, 8)
\Line(  0,72)(20,72)
\Line(  0,75)(20,75)
\Line(  0,78)(20,78)
\Line(20, 7)(60,-2)
\Line(20, 4)(60,-5)
\Line(20,77)(60,86)
\Line(20,74)(60,83)
\ArrowLine(20, 7)(40,40)
\ArrowLine(40,40)(20,73)
\Photon(40,40)(80,40){2}{5}
\Photon(80,40)(115,65){2}{5}
\Photon(80,40)(115,40){2}{5}
\Photon(80,40)(115,15){2}{5}
\GCirc(80,40){10}{0.5}
\Vertex(40,40){2}
\GOval(20, 5)(10,5)(0){.5}
\GOval(20,75)(10,5)(0){.5}
\put(120,62){${V}$}
\put(120,34){${V}$}
\put(120, 6){${V}$}
\put(-20,72){$\Pp$}
\put(-20, 0){$\Pp$}
\end{picture}
}}
\vspace*{-.5em}
\caption{Typical diagram structures for EW di-boson production (left),
EW vector-boson scattering (middle), and tri-boson production (right) in $\Pp\Pp$
collisions, where $VV$ and $VVV$
stand for any combination of two or three W/Z bosons.}
\label{fig:VV-VBS-VVV-diagrams}
\end{figure}

Apart from radiative corrections of the strong interactions described by QCD,
electroweak (EW) radiative corrections play an integral role in predictions 
that should typically aim at percent precision for cross sections or even better
for cross-section ratios or dedicated high-precision observables, such as the
W-boson mass or the effective weak mixing angle at the Z~pole.
In this article, we briefly review the salient features of EW corrections and issues as 
well as corresponding solutions in their calculation. 
Since subtleties related to EW corrections become most apparent in the production
of W and Z bosons, we take production processes
of several EW gauge bosons as illustrating examples,
viz.\ massive di-boson production, vector-boson scattering,
and massive tri-boson production -- processes particularly sensitive to
gauge-boson self-interactions, as illustrated
in Fig.~\ref{fig:VV-VBS-VVV-diagrams}. 
Although suppressed in the figure,
the instability of the produced W/Z~bosons entails serious complications in 
precision calculations, so that W and Z bosons have to be treated as resonant intermediate
states, and all diagrams (with or without resonances) with the same signature of
decay products have to be taken into account.
\looseness-1

\section{Electroweak corrections to scattering cross sections -- concepts, techniques, subtleties}
\label{se:EWRCs}

Although the theories of the strong and EW interactions are both based
on the gauge principle, in these theories 
higher-order corrections and their calculation feature very different
properties. This is due to colour confinement and the necessity to define infrared(IR)-safe
observables in perturbative QCD, the different size of the respective coupling constants,
the massiveness and instability of the weak gauge bosons W/Z, electroweak symmetry
breaking and the existence of the Higgs boson, etc.
In the following we concentrate on EW corrections only and refer to 
review articles~\cite{Denner:2019vbn,Amoroso:2020lgh,Heinrich:2020ybq,%
Tricoli:2020uxr,Covarelli:2021gyz,Maltoni:2022bqs,Jones:2023uzh,Huss:2025nlt}
for the more complete picture including QCD corrections, 
phenomenological applications, and recent developments.
Our conventions and notation consistently follow Ref.~\cite{Denner:2019vbn},
where more details can be found.

\subsection*{\bf Electroweak input parameter schemes} 

The bosonic part of the EW sector on the SM comprises four free parameters:
the two gauge couplings $g_1$, $g_2$ of the gauge group SU(2)$\times$U(1)
and the two parameters $\mu^2,\lambda$ in the Higgs potential.
To make closer contact to measurements, these free parameters are often
traded for the gauge-boson masses $\MW$, $\MZ$, the Higgs-boson mass $\MH$,
and the electromagnetic coupling $\alpha=e^2/(4\pi)$, where $e$ is the electric
unit charge. This fixes the weak mixing angle $\theta_\Pw$ according to
$\sw^2\equiv\sin^2\theta_\Pw=1-\MW^2/\MZ^2$.

The precise phenomenological meaning of the input parameters is fixed by 
imposing renormalization conditions for each free parameter in the course of EW 
renormalization. The masses are typically renormalized via 
{\it on-shell (OS)} conditions, which tie the mass values to the locations of
the respective propagator poles, or via $\MSbar$ prescriptions. 
The value of $\alpha$ can be fixed in the low-energy Thomson limit,
identifying $\alpha$ with the fine-structure constant $\alpha(0)=137.0\dots$, 
at some higher momentum transfer $Q^2$, 
e.g.\ $\alpha(\MZ^2)\approx1/129$ at $Q^2=\MZ^2$,
or through the Fermi constant $\GF$, which is tied to the measured muon
lifetime and leads to $\alpha_{\GF}=\sqrt{2}\GF\MW^2(1-\MW^2/\MZ^2)/\pi\approx1/132$.
The numerical difference of $2{-}6\%$ in these values for $\alpha$ are compensated
in predictions by universal higher-order corrections 
induced by the renormalization of $\alpha$ and the {\it $\rho$-parameter}
for the case of $\alpha_{\GF}$.
Depending on the considered observable or process,
these universal corrections can be absorbed into leading-order (LO) predictions,
in order to minimize the remaining corrections:
\looseness-1

\begin{MYitemize}
\item
{\it $\alpha(0)$-scheme} with input $(\alpha(0),\MW,\MZ)$:
\\
EW corrections to most high-energy processes (GeV range and above)
involve effects of the running $\alpha(Q^2) = \alpha(0)/[1-\Delta\alpha(Q^2)]$,
which includes non-perturbative contributions via $\Delta\alpha(\MZ^2)$ from the
hadronic resonances between $Q^2=0$ and $Q^2=\MZ^2$ that have to be taken from
experiment.
For each ``high-energy coupling factor''~$e$ a correction $\Delta\alpha(\MZ^2)$
appears; for couplings to external on-shell photons no such contribution arises.
\item
{\it $\alpha(\MZ^2)$-scheme} with input $(\alpha(\MZ^2),\MW,\MZ)$:
\\
All corrections $\Delta\alpha(\MZ^2)$ are absorbed into the LO coupling $\alpha(\MZ^2)$
for ``high-energy couplings''~$e$, but in turn new $\Delta\alpha(\MZ^2)$ terms are 
created if~$e$ for an on-shell photon is derived from $\alpha(\MZ^2)$.
\item
{\it $\GF$-scheme} with input $(\alpha_{\GF},\MW,\MZ)$:
\\
In terms of $\Delta\alpha(\MZ^2)$ corrections, this scheme behaves like the 
$\alpha(\MZ^2)$-scheme, but according to $\alpha_{\GF}/\sw^2=\sqrt{2}\GF\MW^2$
it absorbs all universal corrections to the weak charged-current coupling
$g_2\equiv e/\sw$ into LO. This includes, in particular, the well-known 
top-quark-mass-enhanced next-to-leading-order (NLO) corrections 
$\propto\GF\Mt^2$ which are related to the $\rho$-parameter.
\item
{\it LEP scheme} with input $(\alpha(0),\GF,\MZ)$:
\\
This scheme stems from a time in which the experimental precision on $\MW$ 
was not yet competitive. At face value, this scheme produces $\MW=79.\dots\GeV$ 
as LO value, which is very far away form reality. Taking this value as input 
would entail very large universal corrections in predictions (and thus large
theoretical uncertainties), which would be impossible to tame near W~resonances. 
This scheme might be rescued by first calculating $\MW$ from muon decay
including radiative corrections to an appropriate level,
and taking the calculated $\MW$ value as input in one of the previously described schemes.%
\footnote{Cross-section calculations then proceed as if $\MW$ were a measured input value
and either $\alpha(0)$ or $\GF$ has to be dropped from the
list of input parameters.}
\end{MYitemize}
It is clear that there are many processes where none of the schemes is able to
absorb all universal corrections resulting from $\Delta\alpha(Q^2)$ and the $\rho$-parameter
into LO couplings. In many cases, however, this can be achieved upon employing
{\it mixed schemes} in which the LO cross section scales like 
$\alpha(0)^m \alpha(\MZ^2)^n \alpha_{\GF}^\ell$ with non-negative powers
$m,n,\ell$.
Note, however, that only gauge-invariant sets of Feynman graphs can be rescaled
without destroying the consistency of predictions. 

In passing, we note that there are also input schemes tailored to specific 
measurements, such as schemes~\cite{Chiesa:2019nqb}
employing the {\it leptonic effective weak mixing angle}
via $\sin^2\theta_{\mathrm{eff}}^\ell$ as free parameter instead of $\MW$,
which are suited for the determination of $\sin^2\theta_{\mathrm{eff}}^\ell$
in single-Z production.
More details on EW input parameter schemes can be found in Ref.~\cite{Denner:2019vbn}.

\subsection*{\bf W- and Z-boson resonances}

The calculation of radiative corrections to particle resonances is rather delicate,
since perturbation theory in fixed orders (based on real mass parameters) necessarily
produces diverging pole contributions instead of Breit--Wigner-like resonance patterns.
Proper resonance patterns only emerge after Dyson summation of (at least the
imaginary parts of) self-energy corrections, shifting the resonance poles of propagators
into the complex $p^2$ plane for 4-momentum transfer~$p$.
This complex pole location $\mu_P^2=M_P^2-\ri M_P\Gamma_P$ 
is an intrinsic (gauge-invariant) property of an unstable particle~$P$,
so that its real and imaginary parts serve as a well-defined basis for the 
{\it pole mass and width} of~$P$, $M_P$ and $\Gamma_P$.%
\footnote{Note that this definition only coincides with the {\it real on-shell masses}
of W and Z~bosons measured at LEP and the Tevatron at NLO, but differs from the
respective values by $27\MeV$ and $34\MeV$, respectively. 
See Ref.~\cite{Denner:2019vbn} for more details.}

The Dyson summation of propagator corrections
necessarily includes some perturbative orders only partially, 
so that crucial properties of scattering amplitudes or cross sections are jeopardized,
because these are typically proven order by order in perturbation theory.
This, in particular, concerns gauge invariance, 
gauge cancellations at high energies, the cancellation of ultraviolet (UV) and 
IR divergences, unitarity, etc.
These problems are most severe for the W/Z resonances, where the resonance widths
$\Gamma_{\PW/\PZ}$ (i.e.\ the W/Z decay widths) are $\sim 2{-}2.5\GeV$ and thus
much larger than the experimental resolution. 
By contrast, the Higgs-boson resonance cannot be resolved directly owing to
$\Gamma_{\PH}\sim 4\MeV$, and the analysis of the top-quark resonance is
limited by hadronization effects or missing kinematical information of the decay products.

Most predictions for LHC processes with W/Z~resonances are available as
resonance expansions or full off-shell calculations, based on the
following concepts (see Ref.~\cite{Denner:2019vbn} for more details and references):
\begin{MYitemize}
\item
{\it Pole scheme} and {\it pole approximation:}
\\
The idea~\cite{Stuart:1991xk,Aeppli:1993rs} of the {\it pole scheme}
is to first isolate the residue $R$ of each
resonance in the considered amplitude ${\cal M}$
and subsequently to introduce a finite decay width $\Gamma_P$
only in the gauge-independent resonant part,
\begin{align}
{\cal M} &{}=
\frac{R(p^2)}{p^2-M_P^2} + N(p^2)
\;=\;
\frac{R(M_P^2)}{p^2-M_P^2}
+ \frac{R(p^2)-R(M_P^2)}{p^2-M_P^2}
+N(p^2)
\nonumber\\
& \rightarrow\,
\frac{\tilde R(\mu_P^2)}{p^2-\mu_P^2}
\;+\; \frac{R(p^2)-R(M_P^2)}{p^2-M_P^2}+\tilde N(p^2).
\label{eq:poleschemegeneric}
\end{align}
Here, $R(p^2)$ summarizes all contributions developing resonant parts
in the $P$~propagator, and $N(p^2)$ the remaining non-resonant contributions.
The replacement dresses the propagator denominator
as well as its gauge-invariant residue
with the contribution $\ri M_P\Gamma_P$ from the finite decay width,
i.e.\ implements the Dyson summation of the relevant propagator corrections.
Since the role of leading and subleading terms changes when approaching the resonance
vicinity, care has to be taken to avoid double counting of corrections,
so that the non-resonant terms in general have to be modified (indicated
by the tilde on $\tilde N$).
If done carefully the prescription~\refeqf{eq:poleschemegeneric}
respects gauge invariance and can be used to make uniform
predictions in resonant and non-resonant phase-space regions.
In practice, this procedure is full of subtleties, as for instance discussed
in Ref.~\cite{Denner:2019vbn}.
An application of the pole scheme to the weak NLO corrections to single-Z
production has, e.g., been carried out in Ref.~\cite{Dittmaier:2009cr}.

Restricting the pole scheme prescription to the inclusion of the leading term in an
expansion about $p^2=\mu_P^2$, defines the {\it pole approximation (PA)}, which
offers a very solid basis for including NLO corrections in the resonance vicinity
in combination with full off-shell LO predictions.
Note, however, that even in PA subtleties exist, such as the appearance of resonant
contributions -- so-called {\it non-factorizable corrections} --
in the last-but-one term of \refeqf{eq:poleschemegeneric} owing 
to the non-commutativity of the on-shell limit $p^2\to M_P^2$ and the
IR-limit of photon exchange and emission.%
\footnote{Details on their calculation and general results can be found in
\citere{Dittmaier:2015bfe}.}
In practice, defining a PA leaves some freedom concerning the
treatment of off-shell contributions, leading to many possible PA variants;
often only virtual corrections are calculated in PA, while
real corrections are based on full matrix elements.
Estimating the error of the PA in the relative EW corrections by 
the impact of off-shell effects on the intrinsic value of ${\cal O}(\alpha)$ 
leads to the uncertainty estimate ${\cal O}(\alpha/\pi\times \Gamma_V/M_V)$
which is $\lsim0.5\%$ for $V=\PW,\PZ$, including some conservative numerical
enhancement factor. For not too high momentum transfer, where off-shell contributions
can become sizeable, this typically is a realistic estimate.
This has been confirmed in processes with up to three resonances, including
single-W/Z production~\cite{Dittmaier:2014qza} in hadronic collisions,
W-pair production in $\Pep\Pem$~\cite{Denner:2005fg}
and $\Pp\Pp$ scattering~\cite{Biedermann:2016guo}, and 
triple-W production~\cite{Dittmaier:2019twg} at the LHC.

The PA concept also provides a solid basis for including corrections beyond NLO,
where full off-shell calculations are beyond present possibilities.
For instance,
recent results on mixed next-to-next-to-leading order (NNLO) QCD--EW corrections of 
${\cal O}(\alphas\alpha)$ to single-W/Z production show very good agreement between 
PA~\cite{Dittmaier:2014qza,Dittmaier:2015rxo,Dittmaier:2024row}
and full off-shell 
results~\cite{Buonocore:2021rxx,Bonciani:2021zzf,Armadillo:2022bgm,Buccioni:2022kgy}.%
\footnote{For W~production, both PA and off-shell results are not yet fully
complete, but cover the dominant contributions.}

\item
{\it Complex-mass scheme (CMS):}
\\
%
The CMS, as introduced in \citeres{Denner:1999gp,Denner:2005fg} at LO and NLO,
respectively, 
consistently identifies the squared mass of each unstable particle~$P$
with the complex value $\mu_P^2$, both in the propagators and in the couplings,
which become complex.
In particular, couplings involve a complex weak mixing angle fixed via
$\sw^2=1-\mu_\PW^2/\mu_\PZ^2$.
Since the imaginary parts in the complex masses correspond to
higher-order corrections, which are just rearranged,
the usual NLO machinery with Feynman rules and 
counterterms works without modification, without any danger of double-counting.
Only the renormalization constants differ from 
the formalism for stable particles.

NLO calculations deliver uniform predictions with NLO accuracy
everywhere in phase space, both in resonant and non-resonant regions.
Since gauge-boson masses are modified only by an analytic continuation,
all (algebraic) relations expressing gauge invariance are respected:
Amplitudes do not depend on the gauge-fixing procedure (and corresponding gauge
parameters) used in the quantization of the theory, and 
all Ward or Slavnov--Taylor identities of the related Green functions
remain valid (guaranteeing unitarity cancellations).
Owing to the rearranged imaginary parts, however, amplitudes fulfill
the Cutkosky cutting relations, which express unitarity, only up to
terms beyond the completely calculated order. 
Since unitarity cancellations are not disturbed, these higher-order terms
do not amplify the usual uncertainty due to missing higher orders.
\looseness-1

\begin{sloppypar}
The CMS is used in most state-of-the-art predictions for many-particle processes
at the LHC and supported by the frequently used tools
\Gosam~\cite{GoSam:2014iqq},
\MGNLO~\cite{Frederix:2018nkq}, 
\NLOX~\cite{Honeywell:2018fcl},
\mbox{\Openloops2}~\cite{Buccioni:2019sur},
and \Recola \cite{Actis:2012qn,Actis:2016mpe}+\Collier~\cite{Denner:2016kdg} 
in NLO EW calculations.
\end{sloppypar}
%
%
\end{MYitemize}
Other schemes to treat resonances in higher-order calculations are often
tailored to specific applications and less frequently used.
Here we just mention {\it Effective Field Theory} aproaches for unstable 
particles~\cite{Beneke:2003xh,Beneke:2004km,Hoang:2004tg},
based on the separation of scales $\Gamma_P\ll M_P$,
and {\it factorization schemes}, which aim at factorizing 
higher-order corrections from the resonance propagators as far as possible.
Variants of factorization schemes have, e.g., been successfully applied
to single-W/Z production~\cite{Dittmaier:2001ay,Dittmaier:2009cr}.

\subsection*{\bf Photonic corrections}

On the EW side, NLO corrections to hadronic collisions
receive contributions from one-loop amplitudes,
from real photon emission, and from 
real corrections involving photonic initial-state splittings 
$\gamma\to q\bar q$, $q\to q\gamma$, etc.
Photonic corrections often lead to pronounced corrections, way larger than the naively
expected size of ${\cal O}(\alpha)\sim1\%$ for NLO EW effects, typically due to kinematical
redistributions and collinear enhancements of emission effects.
This special role of photonic corrections creates the desire to isolate them and to
include them in perturbative orders that go beyond the one chosen for the remaining EW
corrections. In this context it should, however, be emphasized that there is in general no
possibility to split the EW corrections into a photonic and genuinely weak part 
in a gauge-invariant way. Processes with neutral-current couplings (photon and Z~exchange)
form an exceptional case, in which at least the purely photonic corrections can be isolated, but
in case of W-boson exchange at LO, this splitting is not possible.
In the latter case, at least the isolation of logarithmically enhanced contributions is
possible.
\looseness-1

In the following we briefly review the various types and the phenomenological
relevance of photonic corrections
(for more details, see e.g.\ Ref.~\cite{Denner:2019vbn} and references therein):
\begin{MYitemize}
\item
{\it Photonic bremsstrahlung:}
\\
Real photon emission represents an integral part of EW corrections that has to
be properly combined with loop effects. This combination is non-trivial, because
both real and virtual corrections involve infrared (IR) divergences that are related
and widely cancel according to the Kinoshita--Lee--Nauenberg (KLN) theorem.
At NLO, very general and flexible techniques for this task exist,
the most popular being 
dipole subtraction for photonic corrections~\cite{Dittmaier:1999mb,Dittmaier:2008md}
(following its predecessor of massless QCD~\cite{Catani:1996vz}),
FKS subtraction~\cite{Frederix:2018nkq} (following its QCD proposal~\cite{Frixione:1995ms}),
and phase-space slicing (see, e.g., Ref.~\cite{Harris:2001sx}).
For a general description and more details, see for instance Ref.~\cite{Denner:2019vbn}.

The most prominent difference between QCD and QED radiation is due to the fact that 
leptons (at least muons)
-- in contrast to QCD partons -- can in principle appear as isolated particles
without being ``dressed'' by surrounding photons. 
For such {\it bare} leptons~$\ell$ EW corrections to
distributions (especially in transverse momenta, invariant masses, etc.) are prone
to receive large corrections $\propto\alpha^n\ln^n(m_\ell/Q)$ at the N$^n$LO level
from collinear photonic final-state radiation (FSR),
where the (physical) lepton mass $m_\ell$ acts as regulator for the collinear singularities.
For radiative tails of W/Z~resonances, the corresponding corrections typically receive
several 10\%, e.g.\ $\sim80\%$ in the case of single-Z production with bare muons in the
final state (see, e.g.\ Ref.~\cite{Dittmaier:2009cr}). Dressing final-state leptons
upon recombining collinear photons with the radiating leptons typically reduces the
collinearly enhanced corrections by a factor of~$\sim2$, but still very sizeable
corrections due to FSR remain.
For $\Pep\Pem$ collisions, logarithmically enhanced initial-state radiation (ISR) effects 
$\propto\alpha^n\ln^n(m_{\Pe}/Q)$ generically dominate EW corrections to all
cross sections, without much possibilities to mitigate them by some kind of
photon dressing. 
In hadronic collisions, mass-singular ISR effects are absorbed into the parton distribution
functions (PDFs), so that no particular enhancement from ISR remains in the EW corrections.
Large FSR and ISR effects have to be included in predictions at least in logarithmic
approximations (see, e.g., 
Refs.~\cite{Bertone:2022ktl,Arbuzov:2024tac} and references therein)
if the target precision is one to a few percent.

\item
{\it Photon-induced channels in hadronic collisions:}
\\
Since the photon plays also the role of a parton inside hadrons, EW corrections
to hadronic cross sections comprise contributions from photons in the initial state.
In detail, each process initiated by $q\bar q$ annihilation at LO receives
NLO contributions from $q\gamma$ and $\bar q\gamma$ initial states. 
Neutral final states involving charged particles, such as lepton pair production,
even receive LO contributions from $\gamma\gamma$ collisions.
Most modern NLO calculations that include EW corrections include photon-initiated
channels as well. As for any other parton in the QCD-improved parton model, also
the photon PDF undergoes a PDF redefinition to absorb singular 
contributions from collinear photon emission off charged partons.

In more detail, the photon PDF consists of two different parts:
elastic and inelastic contributions.
While the elastic contributions correspond to the situation where the
hadron stays intact, the inelastic contributions describe collisions in which
the hadron breaks up. Actually, only the inelastic part of the photon PDF
takes part in the mutual transformation of the PDFs induced by splitting processes,
while initial-state photons of elastic collisions are truly on their mass shell
at $Q^2=0$. Note that this difference also has an influence on the choice
of the appropriate input parameter scheme for $\alpha$.
In practice, however, it is not yet standard to split photon-induced processes
into elastic and inelastic components, because this does not yet matter in
predictions at the current level of precision for LHC processes.

Phenomenologically there is hardly any significant impact of 
$q\gamma/\bar q\gamma$-induced EW corrections if 
$q\Pg/\bar q\Pg$-induced QCD corrections involve the same type of diagrams 
with gluons instead of photons. 
Sizeable $q\gamma/\bar q\gamma$-induced corrections mostly appear in processes
involving intermediate W~bosons, which do not couple to gluons. 
See, for instance, the NLO results on WWW production in Sect.~\ref{se:VVV}.
\end{MYitemize}

\subsection*{\bf Photon--jet separation}

NLO predictions for particle particle processes with at least one photon or jet in
the required particle signature involve final states with both photons and jets.
For the corresponding analysis a proper separation between photons and jets is
required, which typically classifies jet-like hadronic showers as photons if the
energy content carried by photons is above some predefined threshold.
Without taking care,
this procedure, however, is prone to break the cancellation of collinear singularities
in the $q^*\to q\gamma$ and $\gamma^*\to q \bar q$ splittings, since the
applicability of the KLN theorem requires inclusiveness
w.r.t.\ massless particles within some cone of jet-like particle clusters.

There are two different approaches in use to implement a separation between nearly
collinear photons and jets that is based on the energy share between photons and hadrons: 
the {\it quark-to-photon fragmentation function}~\cite{Glover:1993xc}
and {\it Frixione isolation}~\cite{Frixione:1998jh}.
The fragmentation function consists of a perturbative part that
cancels residual collinear singularities and of a non-perturbative part that is 
fixed by phenomenological input from dedicated measurements.
Frixione isolation, on the other hand,
defines isolated hard photons by a ``geometrical condition''
that only soft partons can become collinear to the photon.
Both approaches involve resolution parameters defining effective cone sizes which can be
qualitatively matched to each other. 
For $\PW\gamma$~\cite{Denner:2014bna} and $\PZ\gamma$~\cite{Denner:2015fca} 
production at the LHC the two photon--jet separation
approaches yield almost identical results after this matching, but it is not possible
to quantitatively carry this plausible observation over to other processes.
\looseness-1

Issues with the $\gamma^*\to q \bar q$ splitting occur if the $q\bar q$ pair 
can pass the event selection as a single jet in the collinear limit, since the
phase-space integral develops a logarithmic singularity there.
This, for instance, happens in NLO predictions for VBS processes where two tagging
jets are required, but real emission corrections involve up to three partons in
the final state.
As long as only the full integral over some collinear cone around the collinear axis
is required, no differential fragmentation function is required but only its
integrated version, which is inclusive w.r.t.\ the energy sharing in the splitting process.
In Ref.~\cite{Denner:2019zfp} it was shown that this inclusive version, dubbed
{\it photon-to-quark conversion function} there, can be obtained 
from a dispersion integral for the $R$~ratio of the cross
sections for $\Pe^+\Pe^-\to\mathrm{hadrons}/\mu^+\mu^-$, which is directly related
the experimentally known quantity
$\Delta\alpha^{(5)}_{\mathrm{had}}$.

\subsection*{\bf Electroweak corrections at high energies}

At high scattering energies $Q\gg\MW$, EW corrections are enhanced by
logarithmic factors such as
$(\alpha/\sw^2)\ln^2(Q^2/\MW^2)$ and $(\alpha/\sw^2)\ln(Q^2/\MW^2)$
at NLO, and powers of those beyond NLO.
These mass-singular corrections originate from soft and/or
collinear exchange or emission of EW gauge bosons.
In QED and QCD the corresponding double logarithms resulting
from the virtual exchange and real emission of photons/gluons 
cancel each other by virtue of the KLN theorem
and the remaining single logarithms can be
absorbed into PDFs and fragmentation functions.
However, since the masses of the W and Z~bosons provide a physical
cutoff and since the EW charges are not confined, the radiated
real W or Z~bosons can be experimentally reconstructed to a large
extent. 
It is, thus, largely a matter of choice to which extent real massive vector-boson radiation
is included in the definition of the experimental signature of a measured process.
Even if W/Z emission were included in the cross-section definition in a maximally inclusive way,
however,
the level of inclusiveness needed for a full cancellation of the singularities
according to the KLN theorem could not be reached, since the initial states 
of the scattering events carry open EW charges.%
\footnote{By some abuse of language this effect is often called {\it Bloch--Nordsieck 
violation}, although the  Bloch--Nordsieck theorem does not apply to
non-Abelian gauge theories.}
Since NLO corrections due to real radiation are always positive
(owing to their own squared matrix elements $|{\cal M}|^2$),
their lack of inclusiveness leaves behind large negative virtual EW corrections,
which typically grow to several 10\% in the TeV range.

The strongest enhancement of EW corrections at high energies results from the 
kinematic {\it Sudakov regime} in which all invariants
$s_{ij}=2k_i\cdot k_j$ for pairs of particles' four-momenta $k_i,k_j$
become large ($s_{ij}\gg\MW^2$).
The structure of EW corrections in this domain has
been investigated in detail
at NLO and beyond by several 
groups~\cite{Fadin:1999bq,Kuhn:1999nn,Ciafaloni:2000df, Hori:2000tm,%
Denner:2000jv,
Denner:2001gw,Melles:2001dh, Beenakker:2001kf, Denner:2003wi, Jantzen:2005xi,Jantzen:2005az,%
Denner:2006jr}.
As described for example in \citeres{Denner:2001gw,Denner:2003wi,Denner:2006jr}, the
leading EW logarithmic corrections, which are enhanced by
large factors $L=\ln(s_{ij}/\MW^2)$, can be divided into an
SU(2)$\times$U(1)-symmetric part, an electromagnetic part, and a
subleading part induced by the mass difference between $\PW$ and
\PZ~bosons. 
The leading (Sudakov) logarithms $\propto(\alpha L^2)^n$
of electromagnetic origin cancel between
virtual and real (soft) bremsstrahlung corrections, so that
the only source of leading logarithms is,
thus, the symmetric EW
part, which can be characterized by comprising \PW~bosons, \PZ~bosons,
and photons of a common mass $\MW$.
These leading EW Sudakov corrections can be obtained to all orders
from the respective NLO result via exponentiation.
The subleading EW high-energy logarithms can be resummed
upon solving corresponding 
{\it infrared evolution equations}~\cite{Kuhn:1999nn} 
or applying {\it EW Soft--Collinear Effective Theory}~\cite{Chiu:2007yn,Chiu:2008vv}.

The detailed knowledge of the tower of EW high-energy
logarithms is important for a deeper understanding of
EW dynamics, but making use of it in predictions is a subject
that deserves care:
\begin{MYitemize}
\item
It is certainly advisable to make use of full NLO EW corrections,
i.e.\ without applying expansions for high energies, whenever
possible for a given process.
A safe assessment of the
quality of logarithmic approximations usually requires the comparison
to full NLO results. 
Predictions based on logarithmic approximations,
which are widely available in automated form
(see, e.g., Refs.~\cite{Chiesa:2013yma,Bothmann:2020sxm,Pagani:2021vyk,Denner:2024yut})
can at least give a qualitative estimate of EW corrections at high energies.

\item
Beyond NLO, the knowledge of higher-order EW logarithms can be very useful
and exploited to improve pure NLO predictions.
However, particular care has to be taken if the EW logarithms
show large cancellations between leading and subleading terms,
as for instance observed in the case of neutral-current fermion--antifermion
scattering~\cite{Jantzen:2005xi}.
In spite of this uncertainty,
the known part of the tower can at least
deliver estimates for the size of missing corrections and be used in
the assessment of theoretical uncertainties.
\item
While the analytical structure of EW corrections was studied in the literature
in great detail for the Sudakov regime, there is only little knowledge
on EW corrections beyond NLO in more general kinematical situations where
not all invariants $s_{ij}$ are large. Note that there are many
cross sections that are in fact not dominated by the Sudakov regime
in the high-energy limit, including all processes that are
dominated by $t$-channel diagrams. For example,
unless specifically designed cuts are applied, reactions like
W-boson pair production via $\Pep\Pem$, $\Pp\Pp$, or $\gamma\gamma$ collisions
are dominated by the {\it Regge limit}, where the Mandelstam variable
$t$ remains small while $s$ gets large.
It often depends on the specific observable which regime
is probed in high-energy tails of kinematical distributions.
While high-transverse-momentum tails are often dominated by the Sudakov regime,
this is not necessarily the case for tails in invariant masses,
so that approximations for EW high-energy corrections based on the Sudakov regime
are not always reliable.
\end{MYitemize}

\section{Di-boson production}
\label{se:VV}

\begin{figure}
\includegraphics[bb= 0 50 540 810,width=.47\textwidth]{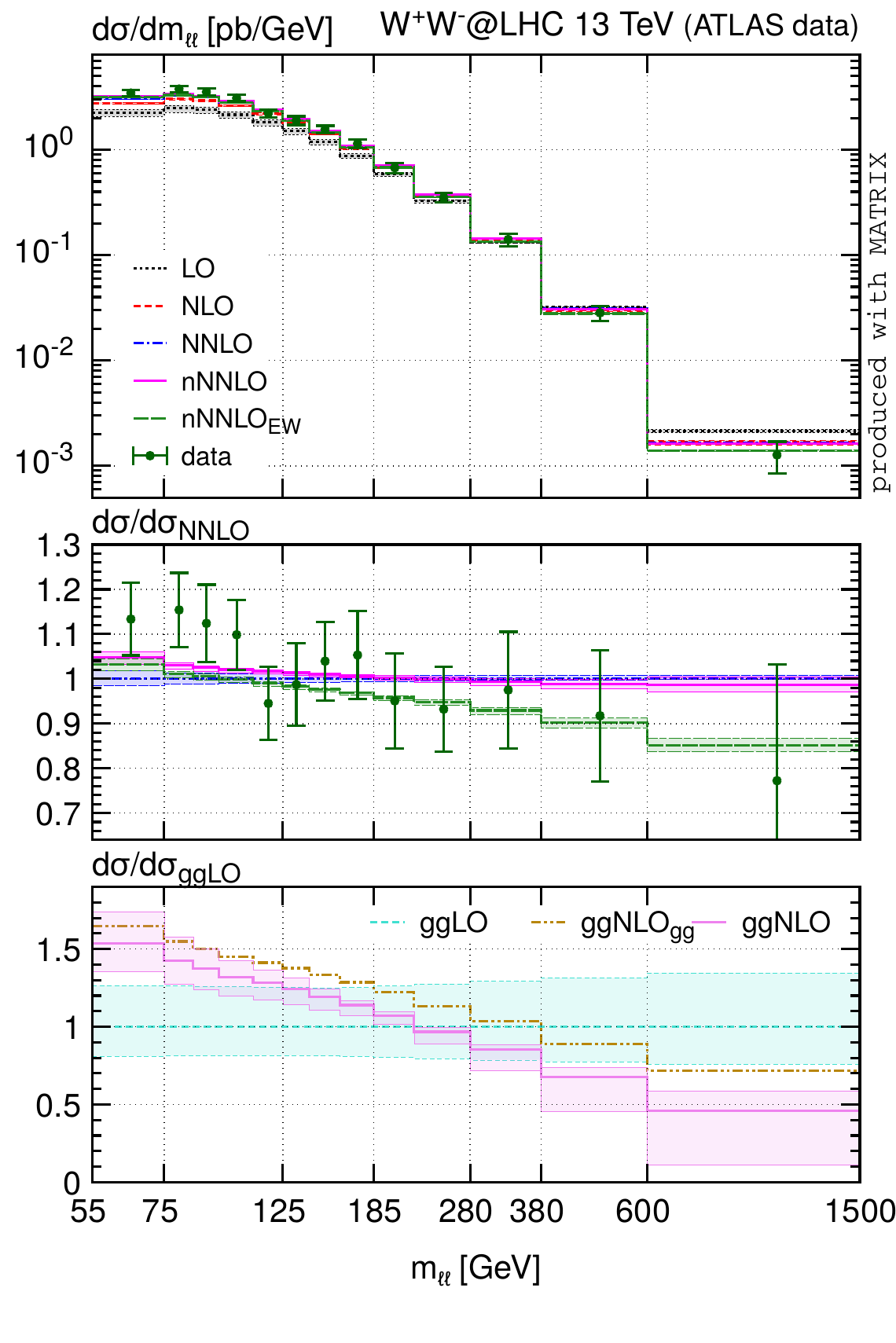}
\raisebox{2em}{\includegraphics[width=.51\textwidth]{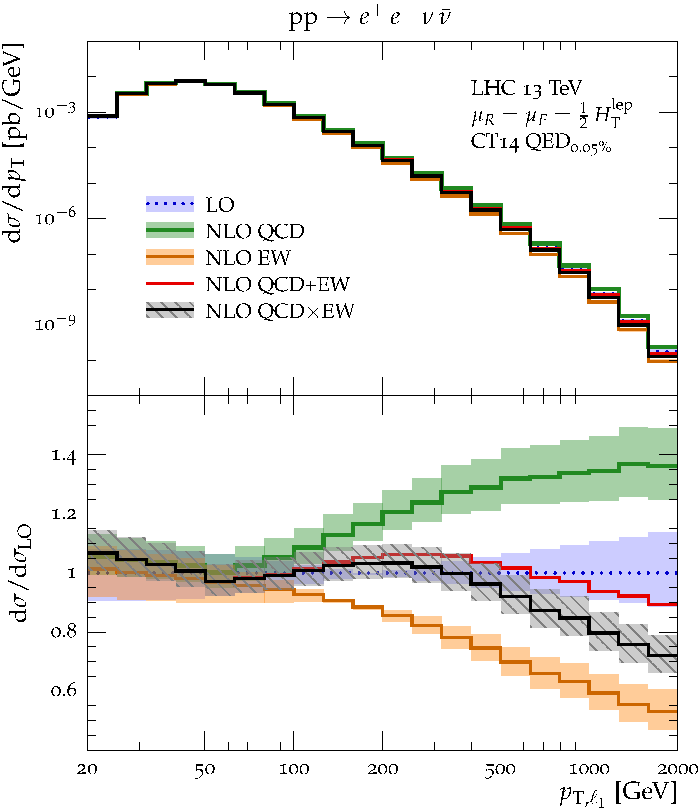}}
\caption{Left: 
state-of-the-art prediction for $\PWp\PWm(\to\Pep\nu_\Pe\mu^-\bar\nu_\mu)$
production provided by \Matrix\ 
(taken from \citere{Grazzini:2020stb}).
Right: 
NLO QCD+EW corrections to $\Pp\Pp\to\PW\PW/\PZ\PZ\to\Pep\Pem\nu\bar\nu+X$
as well as different variants (additive versus multiplicative) to combine
QCD and EW corrections
(taken from \citere{Kallweit:2017khh}).
}
\label{fig:diboson-results}
\end{figure}

Predictions for di-boson production processes at hadron colliders 
have been continuously improved over several decades and have reached a very high 
level of precision, both on the QCD and EW sides.
The current state-of-the-art predictions are provided by the parton-level
Monte Carlo program \Matrix~\cite{Grazzini:2017mhc}, which combines
fixed-order NNLO QCD, NLO EW corrections, and gluon-fusion channels at NLO QCD
with QCD resummations for
single-W/Z production, di-boson production 
(see \citeres{Grazzini:2020stb,Grazzini:2019jkl} and references therein),
and some other process classes.
On its l.h.s., Figure~\ref{fig:diboson-results} illustrates the impact of the
various subcontributions to the prediction for the lepton-pair invariant-mass 
distribution for $\PW\PW$ production in the fully leptonic decay channel,
as well as its confrontation
with experimental data from the ATLAS experiment.
The overall agreement between theory and data is quite good, and the error bars indicate
that upcoming data from LHC Run~3 and the high-luminosity LHC will probe the SM prediction at the
level of EW corrections (shown as nNNLO$_{\mathrm{EW}}$, see \citere{Grazzini:2020stb} for
details).

Here we concentrate on the EW corrections to massive di-boson production, 
which have been calculated for
$\PW\PW$~\cite{Biedermann:2016guo,Kallweit:2017khh},
$\PW\PZ$~\cite{Biedermann:2017oae},
and $\PZ\PZ$~\cite{Biedermann:2016yvs,Biedermann:2016lvg} 
production processes including W/Z~decays and off-shell effects.
Figure~\ref{fig:diboson-results} illustrates the EW corrections to two different
distributions for two slightly different leptonic final states of $\PW\PW$ production; 
the process on the r.h.s.\ also receives contributions from intermediate Z~pairs.
Both examples confirm the enhancement of EW corrections to some 10\% for momentum transfers in
the TeV range. As expected, this enhancement is larger for the transverse-momentum
distribution on the r.h.s., where the high-energy tail is dominated by the Sudakov regime
of the underlying $q\bar q\to\PW\PW$ process,
while the EW corrections to the invariant-mass distribution on the l.h.s.\ are more
moderate. The r.h.s.\ also demonstrates the issue of combining QCD and EW corrections when both
types of corrections get large. The curves labeled QCD$+$EW and QCD$\times$EW compare the
two variants in which the relative corrections $\delta_{\QCD/\EW}$ are combined according to
$1+\delta_{\QCD}+\delta_{\EW}$ and $(1+\delta_{\QCD})(1+\delta_{\EW})$, respectively.
Owing to the factorization property of the EW high-energy logarithms in combination with QCD
effects the multiplicative approach will certainly be closer to the true result, but the
remaining uncertainty from missing non-factorizable QCD--EW corrections should be estimated
carefully.

\begin{figure}
\includegraphics[bb=75 350 370 750,width=0.48\textwidth]{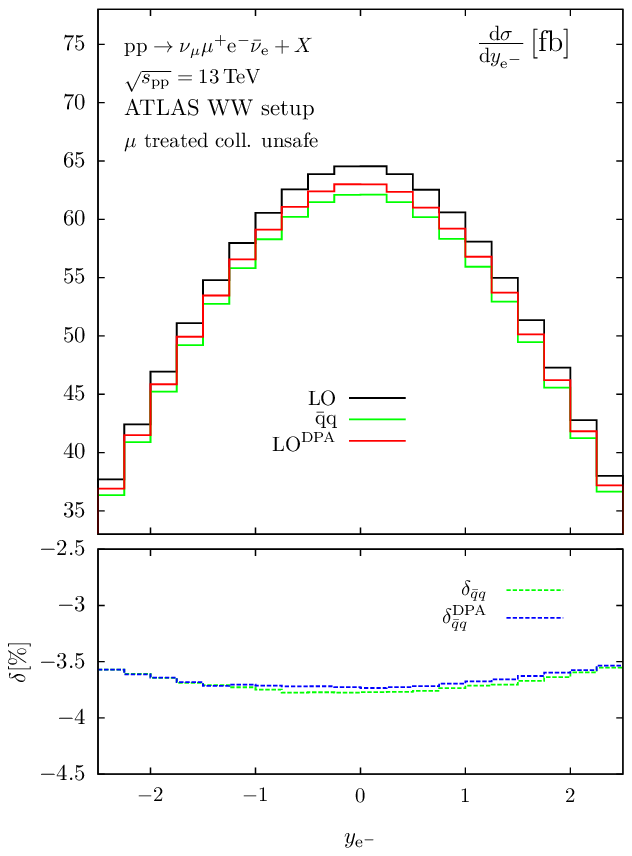} \hfill
\includegraphics[bb=75 350 370 750,width=0.48\textwidth]{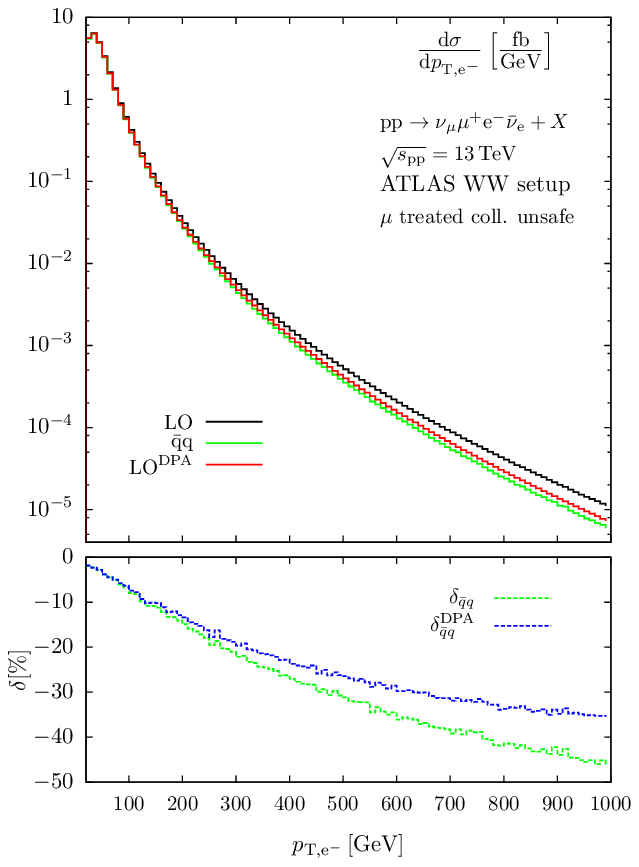}
\caption{Comparison of the NLO EW corrections for $\PW\PW$ production in the
fully leptonic final state to the corresponding DPA,
differential in the rapidity $y_{\Pe^-}$ (left)
and the transverse momentum $p_{\rT,\Pe^-}$ (right) of a charged lepton
(taken from \citere{Biedermann:2016guo}).}
\label{fig:WW-DPA-results}
\end{figure}

For $\PW\PW$~\cite{Billoni:2013aba,Biedermann:2016guo},
$\PW\PZ$~\cite{Le:2022lrp,Le:2022ppa},
and $\PZ\PZ$~\cite{Denner:2021csi}
production, NLO predictions with the W/Z resonances described in 
{\it double-pole approximation (DPA)} are available as well.
Figure~\ref{fig:WW-DPA-results} compares a DPA 
(with real corrections based on full matrix elements)
for the EW corrections to W-pair production
in the leptonic W-decay channel with the corresponding full off-shell results for the distributions
of a charged lepton in its rapidity $y$ and transverse momentum $p_{\rT}$.
Typical for integrated quantities and distributions that do not resolve very high momentum transfer,
the $y$~distribution shows that the DPA reproduces the full result within $\lsim0.5\%$.
On the other hand, the $p_{\rT}$ distribution shows that this approximative quality degrades
in regions where the momentum transfer reaches several $100\GeV$. The figure also illustrates
that the degradation of the DPA results from an upcoming significant impact of
LO contributions that do not feature the double resonance of W~pairs.
Knowing the LO difference $(\LO-\LO^{\DPA})$, the uncertainty of the DPA for the EW corrections
can be realistically estimated without knowing the full off-shell result in NLO
by the product of $(\LO-\LO^{\DPA})$ with the (non-universal) EW corrections in DPA.%
\footnote{For a thorough discussion of such an estimate, see for instance
\citere{Dittmaier:2019twg} for triple-W production and results shown in Sect.~\ref{se:VVV}.}

\section{Scattering of electroweak vector bosons}

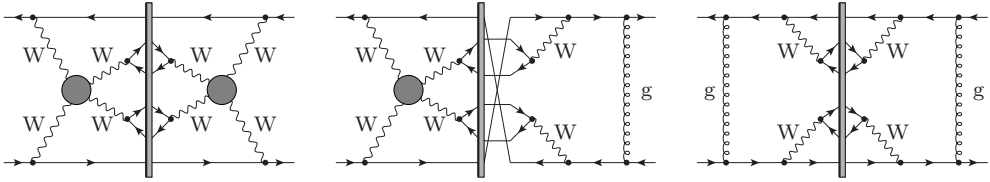
\begin{figure}
{\scriptsize
\setlength{\unitlength}{.55pt}
\begin{picture}(220,130)(0,0)
\SetScale{0.55}
\ArrowLine( 10, 15)( 30, 15)
\ArrowLine( 30, 15)(110, 15)
\ArrowLine(110,115)( 30,115)
\ArrowLine( 30,115)( 10,115)
\ArrowLine(110, 30)( 95, 45)
\ArrowLine( 95, 45)(110, 55)
\ArrowLine(110, 75)( 95, 85)
\ArrowLine( 95, 85)(110,100)
\Photon(30, 15)(60, 65){2}{7}
\Photon(30,115)(60, 65){2}{7}
\Photon(95, 45)(60, 65){2}{6}
\Photon(95, 85)(60, 65){2}{6}
\Vertex(30, 15){2.0}
\Vertex(30,115){2.0}
\Vertex(95, 85){2.0}
\Vertex(95, 45){2.0}
\GCirc(60, 65){10}{.5}
\put(23,84){$\PW$}
\put(23,37){$\PW$}
\put(70,84){$\PW$}
\put(70,37){$\PW$}
\ArrowLine(190, 15)(210, 15)
\ArrowLine(110, 15)(190, 15)
\ArrowLine(190,115)(110,115)
\ArrowLine(210,115)(190,115)
\ArrowLine(125, 45)(110, 30)
\ArrowLine(110, 55)(125, 45)
\ArrowLine(125, 85)(110, 75)
\ArrowLine(110,100)(125, 85)
\Photon(190, 15)(160, 65){2}{7}
\Photon(190,115)(160, 65){2}{7}
\Photon(125, 45)(160, 65){2}{6}
\Photon(125, 85)(160, 65){2}{6}
\Vertex(190, 15){2.0}
\Vertex(190,115){2.0}
\Vertex(125, 85){2.0}
\Vertex(125, 45){2.0}
\GCirc(160, 65){10}{.5}
\put(182,84){$\PW$}
\put(182,37){$\PW$}
\put(138,84){$\PW$}
\put(138,37){$\PW$}
\SetWidth{0}
\GBoxc(110,65)(4,120){.7}
\SetWidth{1.5}
\Line(108,5)(108,125)
\end{picture}
\hfill
\begin{picture}(240,130)(0,0)
\SetScale{0.55}
\ArrowLine( 10, 15)( 30, 15)
\ArrowLine( 30, 15)(110, 15)
\ArrowLine(110,115)( 30,115)
\ArrowLine( 30,115)( 10,115)
\ArrowLine(110, 30)( 95, 45)
\ArrowLine( 95, 45)(110, 55)
\ArrowLine(110, 75)( 95, 85)
\ArrowLine( 95, 85)(110,100)
\Photon(30, 15)(60, 65){2}{7}
\Photon(30,115)(60, 65){2}{7}
\Photon(95, 45)(60, 65){2}{6}
\Photon(95, 85)(60, 65){2}{6}
\Vertex(30, 15){2.0}
\Vertex(30,115){2.0}
\Vertex(95, 85){2.0}
\Vertex(95, 45){2.0}
\GCirc(60, 65){10}{.5}
\put(23,84){$\PW$}
\put(23,37){$\PW$}
\put(70,84){$\PW$}
\put(70,37){$\PW$}
\ArrowLine(230, 15)(210, 15)
\ArrowLine(210, 15)(170, 15)
\ArrowLine(170, 15)(130, 15)
\ArrowLine(170,115)(210,115)
\ArrowLine(130,115)(170,115)
\ArrowLine(210,115)(230,115)
\ArrowLine(145, 45)(130, 30)
\ArrowLine(130, 55)(145, 45)
\ArrowLine(145, 85)(130, 75)
\ArrowLine(130,100)(145, 85)
\Gluon(210, 15)(210,115){2}{14}
\Vertex(210, 15){2.0}
\Vertex(210,115){2.0}
\SetColor{Black}
\Photon(145, 45)(170, 15){2}{6}
\Photon(145, 85)(170,115){2}{6}
\Vertex(170, 15){2.0}
\Vertex(170,115){2.0}
\Vertex(145, 85){2.0}
\Vertex(145, 45){2.0}
\put(220,61){{$\Pg$}}
\put(160,87){$\PW$}
\put(160,31){$\PW$}
\Line(112, 15)(130,115)
\Line(112,115)(130, 15)
\Line(130, 30)(110, 30)
\Line(110, 55)(130, 55)
\Line(130, 75)(110, 75)
\Line(110,100)(130,100)
\SetWidth{0}
\GBoxc(110,65)(4,120){.7}
\SetWidth{1.5}
\Line(108,5)(108,125)
\end{picture}
\hfill
\begin{picture}(220,130)(0,0)
\SetScale{0.55}
\ArrowLine( 10, 15)( 30, 15)
\ArrowLine( 30, 15)( 70, 15)
\ArrowLine( 70, 15)(110, 15)
\ArrowLine( 70,115)( 30,115)
\ArrowLine(110,115)( 70,115)
\ArrowLine( 30,115)( 10,115)
\ArrowLine(110, 30)( 95, 45)
\ArrowLine( 95, 45)(110, 55)
\ArrowLine(110, 75)( 95, 85)
\ArrowLine( 95, 85)(110,100)
\Gluon(30, 15)(30,115){2}{14}
\Vertex(30, 15){2.0}
\Vertex(30,115){2.0}
\SetColor{Black}
\Photon(95, 45)(70, 15){2}{6}
\Photon(95, 85)(70,115){2}{6}
\Vertex(70, 15){2.0}
\Vertex(70,115){2.0}
\Vertex(95, 85){2.0}
\Vertex(95, 45){2.0}
\put(15,61){{$\Pg$}}
\put(65,87){$\PW$}
\put(65,31){$\PW$}
\ArrowLine(190, 15)(210, 15)
\ArrowLine(150, 15)(190, 15)
\ArrowLine(110, 15)(150, 15)
\ArrowLine(190,115)(150,115)
\ArrowLine(150,115)(110,115)
\ArrowLine(210,115)(190,115)
\ArrowLine(125, 45)(110, 30)
\ArrowLine(110, 55)(125, 45)
\ArrowLine(125, 85)(110, 75)
\ArrowLine(110,100)(125, 85)
\Gluon(190, 15)(190,115){2}{14}
\Vertex(190, 15){2.0}
\Vertex(190,115){2.0}
\SetColor{Black}
\Photon(125, 45)(150, 15){2}{6}
\Photon(125, 85)(150,115){2}{6}
\Vertex(150, 15){2.0}
\Vertex(150,115){2.0}
\Vertex(125, 85){2.0}
\Vertex(125, 45){2.0}
\put(200,61){{$\Pg$}}
\put(140,87){$\PW$}
\put(140,31){$\PW$}
\SetWidth{0}
\GBoxc(110,65)(4,120){.7}
\SetWidth{1.5}
\Line(108,5)(108,125)
\end{picture}
}
\vspace*{-.5em}
\caption{Typical LO interference diagrams for VBS processes, contributing to the LO
cross sections at 
${\cal O}(\alpha^6)$, ${\cal O}(\alphas^2\alpha^4)$, and ${\cal O}(\alphas\alpha^5)$,
respectively.}
\label{fig:LOVBSdiags}
\end{figure}

EW vector-boson scattering (VBS), $VV\to VV$, is mostly analyzed via final states with
leptonically decaying EW gauge bosons, so that the experimental signature comprises
four leptons and two jets ($4\ell2\Pj$).
As illustrated in Fig.~\ref{fig:LOVBSdiags},
the contributions to the LO cross section for $4\ell2\Pj$ production scales like 
${\cal O}(\alpha^6)$, ${\cal O}(\alphas^2\alpha^4)$, and ${\cal O}(\alphas\alpha^5)$
for the purely EW graphs, squared graphs with initial-state gluons or gluon exchange, and
the interferences between EW and gluon-exchange graphs, respectively.
Apart from the like-sign VBS channel with $\PW^\pm\PW^\pm$ intermediate states,
the respective final states with four leptons and two quarks receive large LO
contributions of ${\cal O}(\alphas^2\alpha^4)$ from gluon fusion.
Note also that all VBS channels receive LO contributions from tri-boson production,
$\Pp\Pp\to VVV+X$.
In order to enhance the purely EW VBS contribution of ${\cal O}(\alpha^6)$ to the
overall cross section, dedicated VBS cuts have been devised, which essentially
demand a pair of forward--backward pointing jets with an invariant mass of
$M_{\Pj\Pj}\gsim500\GeV$ and a large rapidity gap of $|\Delta y_{\Pj\Pj}|\gsim2.5$.
These cuts strongly suppress the tri-boson background 
and enhance the EW VBS contribution to like-sign W~scattering to about $90\%$,
but the QCD part in the other VBS channels still remains about $80\%$.

\begin{figure}
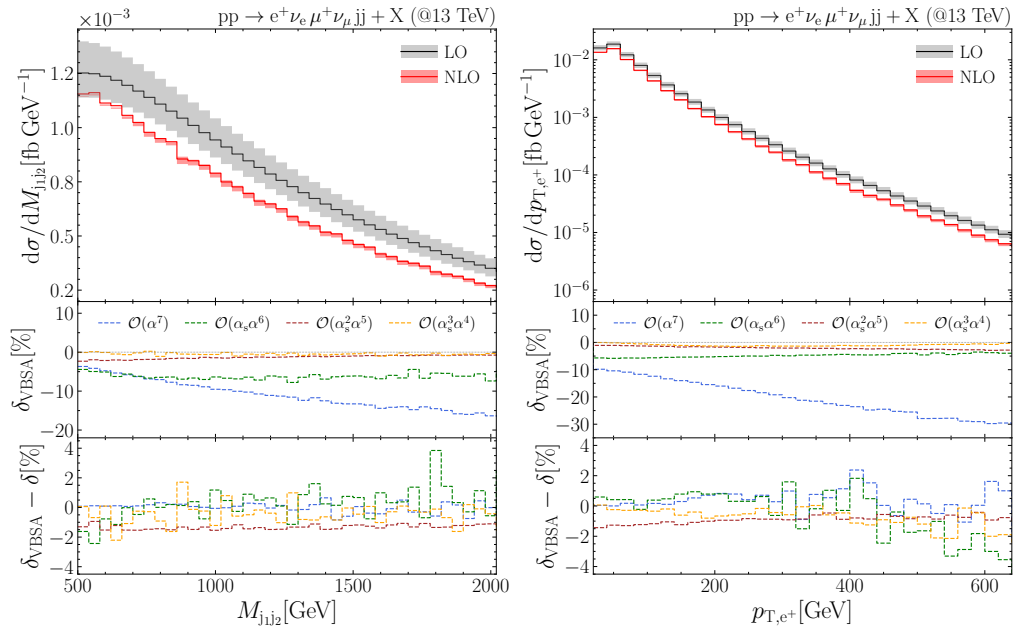

\includegraphics[bb= 10 6 458 557,clip,page=7,width=0.49\textwidth]{nlos_vs_vbsa} \hfill
\includegraphics[bb= 10 6 458 557,clip,page=8,width=0.49\textwidth]{nlos_vs_vbsa}
\vspace*{-.5em}
\caption{NLO predictions for like-sign W-pair production at the LHC, differential
in the di-jet invariant mass (left) and the transverse momentum
of a charged lepton  (right). In the lower panels,
the full relative corrections $\delta$ of the indicated perturbative order are compared
with their counterparts in VBSA. (Taken from \citere{Dittmaier:2023nac}.)}
\label{fig:VBSresults}
\end{figure}

The NLO corrections to the cross section comprise contributions of the orders
${\cal O}(\alpha^7)$, ${\cal O}(\alphas\alpha^6)$,
${\cal O}(\alphas^2\alpha^5)$, and ${\cal O}(\alphas^3\alpha^4)$.
While the QCD corrections to the EW and QCD channels are known already for about 20~years
for all relevant VBS channels,
the NLO towers including EW corrections 
(up to very few remaining, less prominent orders)
have been completed in recent years for 
$\PW^\pm\PW^\pm$~\cite{Biedermann:2016yds,Biedermann:2017bss,Dittmaier:2023nac},
$\PW\PZ$~\cite{Denner:2019tmn},
$\PZ\PZ$~\cite{Denner:2020zit,Denner:2021hsa}, and
$\PW^\pm\PW^\mp$~\cite{Denner:2022pwc} scattering.
The NLO corrections to all VBS channels share the feature that the purely EW corrections,
i.e.\ the ones of ${\cal O}(\alpha^7)$, are quite large, typically $\sim-15\%$
normalized to the EW LO contribution that actually features EW VBS.
All other NLO orders are less prominent, see Fig.~\ref{fig:VBSresults}.
In Ref.~\cite{Biedermann:2016yds} this large effect was traced back to the impact 
of EW Sudakov corrections $\propto(\alpha/\sw^2)\ln^2(Q^2/\MW^2)$ plus subleading
single-logarithmic universal corrections, where $Q\sim400\GeV$ corresponds to the
mean invariant mass of the produced four-lepton system.

The enormous complexity of NLO calculation for $2\to6$ particle processes like VBS
calls for proper approximations, e.g.\ to render predictions in SM extensions less costly.
In this spirit, early calculations of QCD corrections to the EW contribution had been 
performed in ``$t$-channel approximation'', which is justified by the strong suppression
of gluon exchange between the two quark lines owing to colour conservation.
In a similar spirit, Ref.~\cite{Dittmaier:2023nac} discusses a ``VBS approximation (VBSA)''
that neglects all tri-boson contributions, gluon-fusion contributions (if there are any),
and corrections to LO interference diagrams between $t$- and $u$-channels. 
Moreover, the virtual corrections treat the
two produced vector-boson resonances in DPA, so that
the computational effort at NLO is enormously reduced w.r.t.\ the full off-shell calculation.
Figure~\ref{fig:VBSresults} illustrates that the VBSA for like-sign W~scattering is good
within $\lsim1.5\%$ in the experimentally relevant phase space.
For extremely high momentum transfer, of course, limitations of the VBSA are expected,
similar to the observations made for di- and tri-boson production 
(cf.~Sects.~\ref{se:VV} and \ref{se:VVV}).
\looseness-1

\section{Tri-boson production}
\label{se:VVV}

\begin{figure}
\includegraphics[page=1,width=0.3\textwidth]{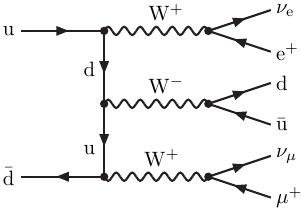} \hfill
\includegraphics[page=2,width=0.3\textwidth]{WWW-fdiagrams.pdf} \hfill
\includegraphics[bb = 235 690 360 775, clip,width=0.3\textwidth]{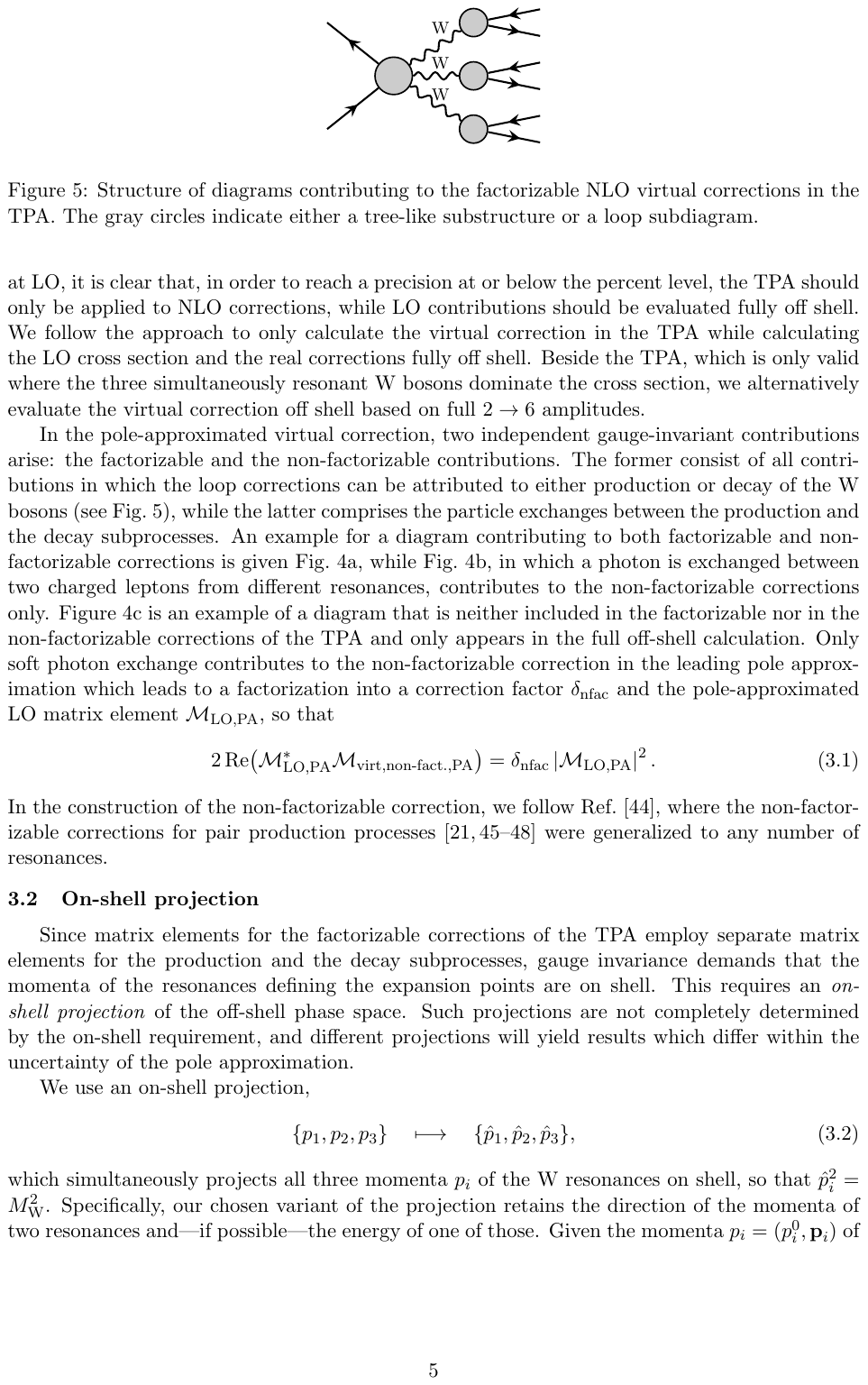} 
\vspace*{-.5em}
\caption{Sample diagrams for $\PW\PW\PW$ production:
LO diagram with three potentially resonant W~bosons (left),
LO diagram contributing to $\PW\PH(\to\PW\PW^*)$ production (middle),
and structural diagram for the triple-pole approximation for WWW~production (right).}
\label{fig:WWW-diags}
\end{figure}

EW tri-boson production is typically analysed in fully leptonic final states ($6\ell$)
or in final states with four leptons and two jets ($4\ell2\Pj$). The former
provides the cleaner experimental signature, but the latter features larger 
cross sections. Note also that the $4\ell2\Pj$ signature is the same as for VBS
processes, i.e.\ the same sets of Feynman diagrams are relevant for predictions, 
but the experimental analyses focus on different kinematical regions.
Tri-boson analyses are carried out without applying VBS cuts, in order to
make the impact of $VVV$ production diagrams, as shown on the l.h.s.\ of
Fig.~\ref{fig:WWW-diags}, more prominent.
As shown in the middle of the figure, some $VVV$ channels ($V\PW\PW$ and $V\PZ\PZ$
production) also receive contributions from $V\PH$ production; in case of
$V\PW\PW$ production the $V\PH$ contribution cannot be suppressed if at least
one of the $\PW$ bosons resulting from the Higgs-boson resonance decays leptonically.
\looseness-1

\begin{figure}
\includegraphics[width=0.5\textwidth]{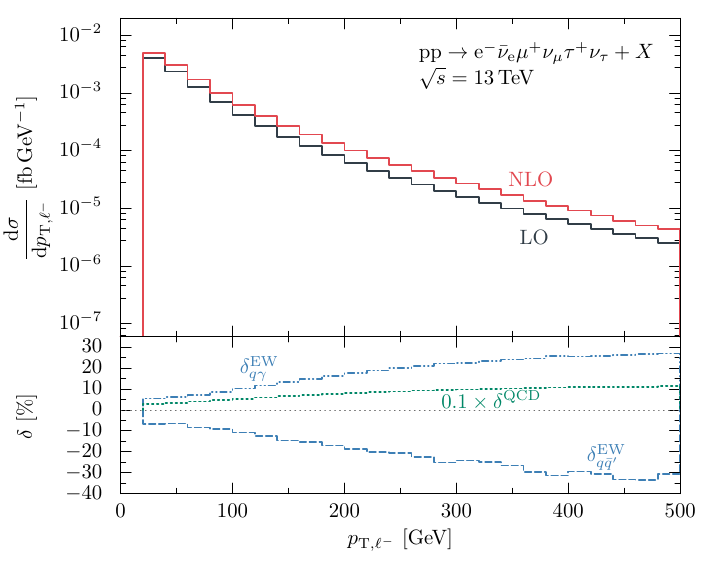}%
\includegraphics[width=0.5\textwidth]{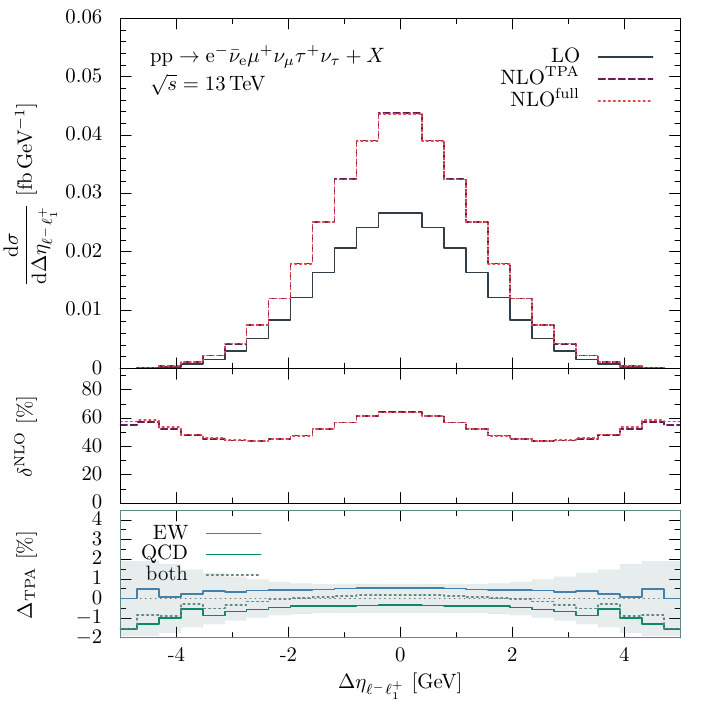}%
\vspace*{-.5em}
\caption{NLO prediction for $\PW^+\PW^+\PW^-$ production in the fully leptonic final state,
differential in the transverse momentum of one of the leptons (left),
and the rapidity difference between the lepton $\ell^-$ and the leading lepton $\ell^+$
(with higher $p_{\rT}$, right). On the left, the relative corrections of different origin
are shown; on the right, the full NLO correction is compared to its counterpart in
TPA.
(Taken from \citere{Dittmaier:2019twg}.)}
\label{fig:WWW-6l-results}
\end{figure}

For fully leptonic final states, the LO $VVV$ cross section scales like
${\cal O}(\alpha^6)$, so that genuine QCD and EW corrections of 
${\cal O}(\alphas\alpha^6)$ and ${\cal O}(\alpha^7)$, respectively exist at NLO.
These corrections have been calculated in Refs.~\cite{Schonherr:2018jva,Dittmaier:2019twg}
for the full off-shell WWW processes. 
Additionally, a triple-pole approximation (TPA) for the NLO corrections based on the
contributions featuring an resonance enhancement by all three intermediate W~bosons
has been worked out and discussed in Ref.~\cite{Dittmaier:2019twg}.
While QCD corrections are about $\sim40\%$, the EW corrections receive contributions
of $\sim\mp(7{-}8)\%$ from $q\bar q$ and $q\gamma$ initial states to integrated cross
sections that tend to cancel against each other. The cancellation, however, is purely
accidental, strongly depends of phase-space cuts, and is less prominent in distributions,
where the EW corrections in the $q\bar q$ channels show the typical Sudakov enhancement 
at high momentum transfer.
Figure~\ref{fig:WWW-6l-results} illustrates this fact for distributions in the
transverse momentum of a charged decay lepton and the rapidity difference of two
charged leptons.
The figure on the r.h.s.\ also demonstrates that both the QCD and EW corrections are very
well approximated by the TPA, typically within $\lsim0.5\%$ in phase-space regions that
dominate the cross section. The gray shaded area indicates an error estimate for the TPA,
nicely covering the difference between full off-shell NLO calculation and TPA.
Similar to the observation made for the DPA in di-boson production, the quality of the TPA
degrades in high-energy tails of distributions (like the missing energy carried away by neutrinos)
that are sensitive to ``collective recoil effects'' induced by contributions with less than
three W-boson resonances (see Ref.~\cite{Dittmaier:2019twg} for more details).
\looseness-1

\begin{figure}
\includegraphics[width=0.5\textwidth]{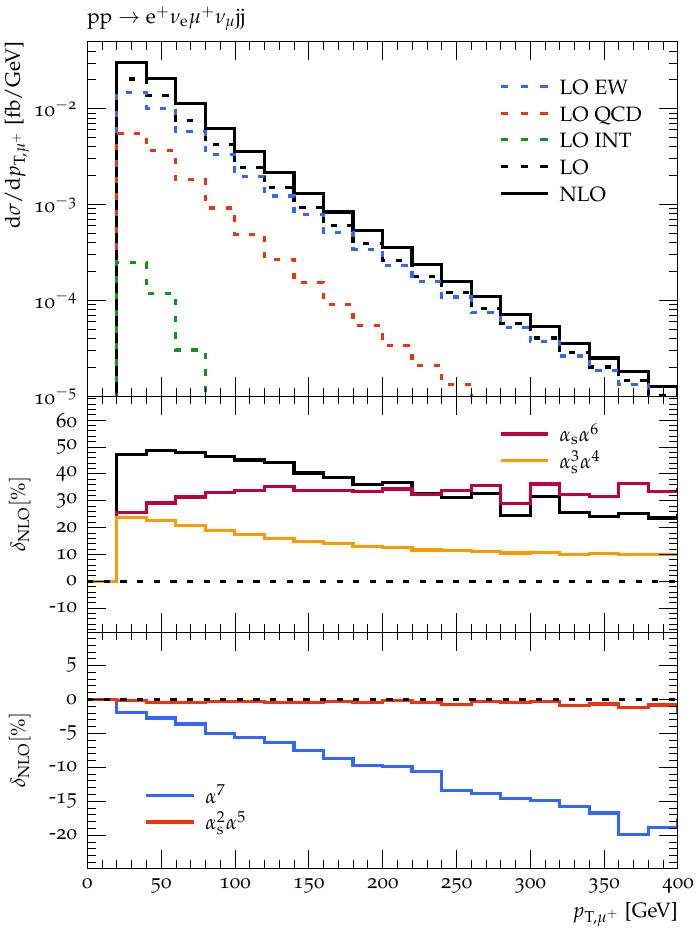}%
\includegraphics[width=0.5\textwidth]{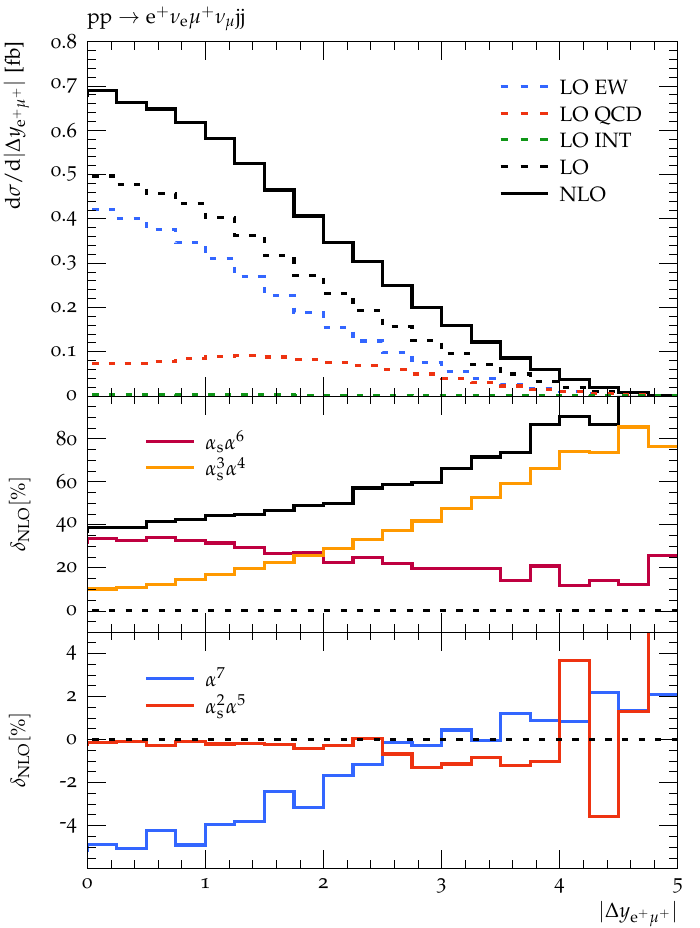}%
\vspace*{-.5em}
\caption{NLO prediction for $\PW\PW\PW$ production with 
one hadronically and two leptonically decaying $\PW$ bosons, differential
in the transverse momentum of one of the leptons (left)
and the rapidity difference of the two charged leptons (right).
(Taken from \citere{Denner:2024ufg}.)}
\label{fig:WWW-4l2j-results}
\end{figure}

For $VVV$ production with a hadronically decaying vector boson, 
the same tower of LO and NLO contributions as for VBS exists.
For some WWW and WWZ/WZZ production channels the NLO towers have been calculated
in Refs.~\cite{Denner:2024ufg} and \cite{Denner:2024ndl}, respectively
(at least partially in the later case).
Some selected results for WWW production are shown in 
Fig.~\ref{fig:WWW-4l2j-results}, revealing crucial differences to the corrections
found for the same final state in the kinematical VBS regime
(cf.~Fig.~\ref{fig:VBSresults}).
Genuine EW corrections of ${\cal O}(\alpha^7)$ to the integrated cross section
are of the order of $\sim-7\%$, similar to the case of WWW production with leptonic
final states and less pronounced than in the VBS case.
The usual EW Sudakov enhancement for high momentum transfer drives these corrections to 
$-(10{-}20)\%$ for transverse lepton momenta of $200{-}400\GeV$; for higher $p_{\rT}$
the cross section is suppressed by two orders of magnitude.
QCD corrections of ${\cal O}(\alphas\alpha^6)$ and ${\cal O}(\alphas^3\alpha^4)$ reach
several 10\%, rendering the inclusion of 
NNLO QCD corrections desirable.%
\footnote{In fact, the SM prediction undershoots LHC data~\cite{ATLAS:2024wla} 
by an amount that could be
accounted for by missing NNLO QCD effects, assuming a similar pattern of (N)NLO QCD corrections
observed in WW production.}
Finally, it should be mentioned that the cross sections for WWW production shown in 
Fig.~\ref{fig:WWW-4l2j-results} contain an irreducible contribution of $\sim40\%$
originating from WH~production, so that the resulting values for the individual
NLO orders actually are a complicated mix of corrections to the genuine WWW,
WH, and VBS subprocesses.

\section{Concluding remarks}

Despite their complexity -- in particular in processes with resonances and many
particles in the final state -- the nature of EW corrections is well understood
at NLO, and powerful tools for an automatic calculation exist.%
\footnote{As experience shows, this is no guarantee for correct results though.}
In this article, we have briefly reviewed salient features and subtleties in the
calculation of EW corrections, such as absorbing large universal corrections in
appropriate input parameter schemes, the gauge-invariant treatment of particle resonances,
and the sources of prominent EW corrections.
As illustrative examples we have picked W/Z~production processes at the LHC
featuring two or three gauge-boson resonances and discussed the emerging EW corrections
as well as the possibility to approximate them via leading-pole approximations.
The good quality achieved by such approximations provides a solid basis for studying
the polarizations of the W/Z~bosons, for including corrections beyond NLO,
and for predictions within complicated
SM extensions, where full off-shell calculations might become too cumbersome.

Current and future frontiers in this context, for instance,
comprise the proper combination of EW and QCD
corrections, which ultimately requires the calculation of NNLO
corrections of the mixed QCD/EW order ${\cal O}(\alphas\alpha)$,
a proper treatment of EW corrections beyond NLO for energies at the TeV scale,
where corrections grow to some 10\% owing to EW Sudakov enhancements,
and a global assault on genuine EW corrections at the NNLO level and beyond, necessary to
account for the ultimate precision at potential future $\Pep\Pem$ colliders.


\bibliographystyle{elsarticle-num_mod}
\bibliography{EWTH-Dittmaier-refs}

\end{document}